\documentclass[journal]{IEEEtran}
\usepackage{cite}
\usepackage{amsmath,amssymb,amsfonts}
\usepackage{tikz}
\usepackage{algorithmic}
\usepackage{algorithm}
\usepackage{graphicx}
\usepackage{subfig}
\usepackage{textcomp}
\usepackage{bbm}
\usepackage{xcolor}
\usepackage{soul}
\usepackage{balance}
\def\BibTeX{{\rm B\kern-.05em{\sc i\kern-.025em b}\kern-.08em
    T\kern-.1667em\lower.7ex\hbox{E}\kern-.125emX}}

\begin{document}

\title{ Online Learning for Autonomous Management of  Intent-based 6G Networks}

\author{Erciyes Karakaya\thanks{This work was supported by The Scientific and Technological Research Council of Turkey (TUBITAK) through the 1515 Frontier Research and Development Laboratories Support Program under Project 5169902, and has been partly funded by the European Commission through the Horizon Europe/JU SNS project Hexa-X-II (Grant Agreement no. 101095759).

Erciyes Karakaya is with the Faculty of Engineering and Natural Sciences, Sabanci University, Turkey. E-mail: rerciyes@sabanciuniv.edu}, {Ozgur Ercetin}\thanks{O.~Ercetin is with the Faculty of Engineering and Natural Sciences, Sabanci University, Turkey. E-mail: oercetin@sabanciuniv.edu},  Huseyin Ozkan\thanks{H.~Ozkan is with the Faculty of Engineering and Natural Sciences, Sabanci University, Turkey. E-mail: huseyin.ozkan@sabanciuniv.edu}, Mehmet Karaca\thanks{M.~Karaca is with Ericsson Research, Turkey. E-mail: mehmet.karaca@ericsson.com}, Elham Dehghan Biyar\thanks{E.D.~Biyar is with Ericsson Research, Turkey. E-mail: elham.dehghan.biyar@ericsson.com}, Alexandros Palaios\thanks{A.~Palaios is with Ericsson Research, Germany. E-mail: alex.palaios@ericsson.com}}%

\maketitle

\begin{abstract}
The growing complexity of networks and the variety of future scenarios with diverse and often stringent performance requirements call for a higher level of automation. Intent-based management emerges as a solution to attain high level of  automation, enabling human operators to solely communicate with the network through high-level intents. The intents consist of the targets  in the form of expectations (i.e., latency expectation) from a service and based on the expectations the required network configurations should be done accordingly. It is almost inevitable that when a network action is taken to fulfill one intent, it can cause negative impacts on the performance of another intent, which results in  a conflict. In this paper, we aim to address the conflict issue and autonomous management of intent-based networking, and propose an online learning method based on the hierarchical multi-armed bandits approach for an effective management.  Thanks to this hierarchical structure, it performs an efficient exploration and exploitation of network configurations with respect to the dynamic network conditions. We show that our algorithm is an effective approach regarding resource  allocation and satisfaction of intent expectations.
\end{abstract}


\section{Introduction}
The expectation from 5G is to support the variety of future scenarios with different type of services such as Ultra Reliable Low Latency Communication (URLLC), and massive Internet of Things (mIoT), Mobile Broadband (eMBB). On the other hand, we expect to see more services with different performance requirements and 6G is expected to be ready to fulfill all these requirements, which highlights that  the importance of autonomous management  of future network is becoming more critical \cite{hx2}. However, autonomous network management has been the holy grail for telecommunication operators for decades. Traditionally, network automation refers to automating the configuration, adding or disconnecting services, deployment, and operating physical and virtual devices to reduce cost, save time, and increase efficiency. This is typically achieved by rule-based mechanisms developed in hindsight for certain events and issues \cite{r20}. However, future novel applications and services will probably require  a proactive approach where the action is to be taken before a disruptive event occurs. With this aim, Intent-based networking (IBN) emerges as a pivotal instrument for the management of networks and services within the forthcoming generation of networks \cite{r10}, \cite{tmforum}, \cite{etsizsm}. Intents carry the high-level expectation of a service such as throughput or latency, and are declarative stating the desired state but not how to achieve it. Achieving such desired intent expectations, autonomous management of intent is crucial as  it is a very difficult task for a human to adjust the required network configuration depending on varying and dynamic intent expectations.

In this regard, another important dimension of an autonomous network lies in the intelligent and cognitive {\em closed-loops} (CLs) capable of learning optimal behavior through interaction with the network while performing multiple functions. We envision that each CL is mapped to an intent expectation and can autonomously attempt to change a network configuration to satisfy its expectation. Nevertheless, because CLs operate independently,  network modification by one CL may impede the performance of other CLs, inevitably leading to conflicts. 

In the current work, we focus on the interaction between  CLs performing different functionalities while achieving certain goals both individually and collectively. In our approach, multiple independent and correlated low level CLs, namely \textit{child } CL,  administer the network to manage its associated intent expectation or Key Performance Indicator (KPI) autonomously \cite{r10}. There is also a hierarchical management with a \textit{parent CL} that has a broader view of the possible issues, responsibilities, and accountability for actions or decisions that need to be taken at the lower levels. For example at a broader perspective, a high level CL (parent) can be able to observe the issues from multiple low level CLs (child). Additionally, higher-level CLs maintain long-term planning and make reasoning. We develop a hierarchical based multi-arm bandit (MAB) framework that handles the operations at child and parent CLs to satisfy the intent requirements by resolving the conflict among the child CLs in a dynamically changing environment.

In \cite{baktir}, the authors aim to address the conflict issue among different CLs by predicting the impact of every action proposed to satisfy the KPI requirement. However, such prediction is not an easy task especially when there are many KPI requirements. In  \cite{perepu}, model-free multi-agent reinforcement learning approach is proposed,  where each agent is responsible for tuning a parameter related to the KPI requirement. In this work, the agents are trained in conjunction with each other to manage multiple services and is not easy to extend the training when a different or chaining environment is faced. In \cite{banerjee} and \cite{cinemre} apply  a game theoretical solutions to resolve the conflict, and however it is not clear the performance  of those solutions when the requirements, priority of KPI, and environments change dynamically over time.  Our proposed approach can adapt to dynamic conditions by only performing the training at the child agent level that can be utilized in  different scenarios.

Our work addresses resource allocation challenges in networks with limited resources, aiming to achieve a mutually agreeable solution for all network services. We recognize that network conditions and services can be highly dynamic, and our system is designed to be adaptable to these potentially non-stationary changes. We adopt a hierarchical multi-armed bandit structure within the limited resource, i.e bandwidth, of partially known system. Our approach aims to optimize critical network service metrics, such as Quality of Experience (QoE) and latency, to improve overall network performance. Ultimately, we aim to create a robust and adaptive network system that efficiently allocates resources and satisfy different intent expectation in an autonomous way, even in dynamically changing environments.

\section{Problem Formulation}

In our model, we consider that there are $N$ services with different intent expectations, and we assume that a network operator makes a contract for each service and Service Level Agreements (SLAs) with the desired intents to be established. The objective of the network operator is to fulfill the intents by taking into account the business value of them. To do that, the operator needs to carefully make decisions to maximize its benefit. One of the main challenge is that the requirements of the intents can change over time such that the KPI expectations can vary, or even a new intent can be introduced or an existing one can be removed from the system. Under such dynamic and unknown conditions, the operator must be adaptive and optimally allocate its resource by taking the necessary network configurations and actions.

There might be a large number of intent requirements, and  one solution may be to assign a single CL to handle all the intent expectations. However, this solution requires an centralized authority, and  will become infeasible as the number of intent increases. Therefore, in this paper,  we assign a CL to each intent expectation (i.e.,KPI) and these CLs autonomously take decisions to meet their KPI expectations. For example, an URLLC service can have an intent that dictates an expectation on end-to-end latency (i.e., latency should be less than 30 ms.), and a video service can have an intent with QoE expectation. Then, two independent CLs are assigned to the latency and QoE expectations.

\begin{figure}[t]
\centerline{\includegraphics[width=\columnwidth,height=48mm]{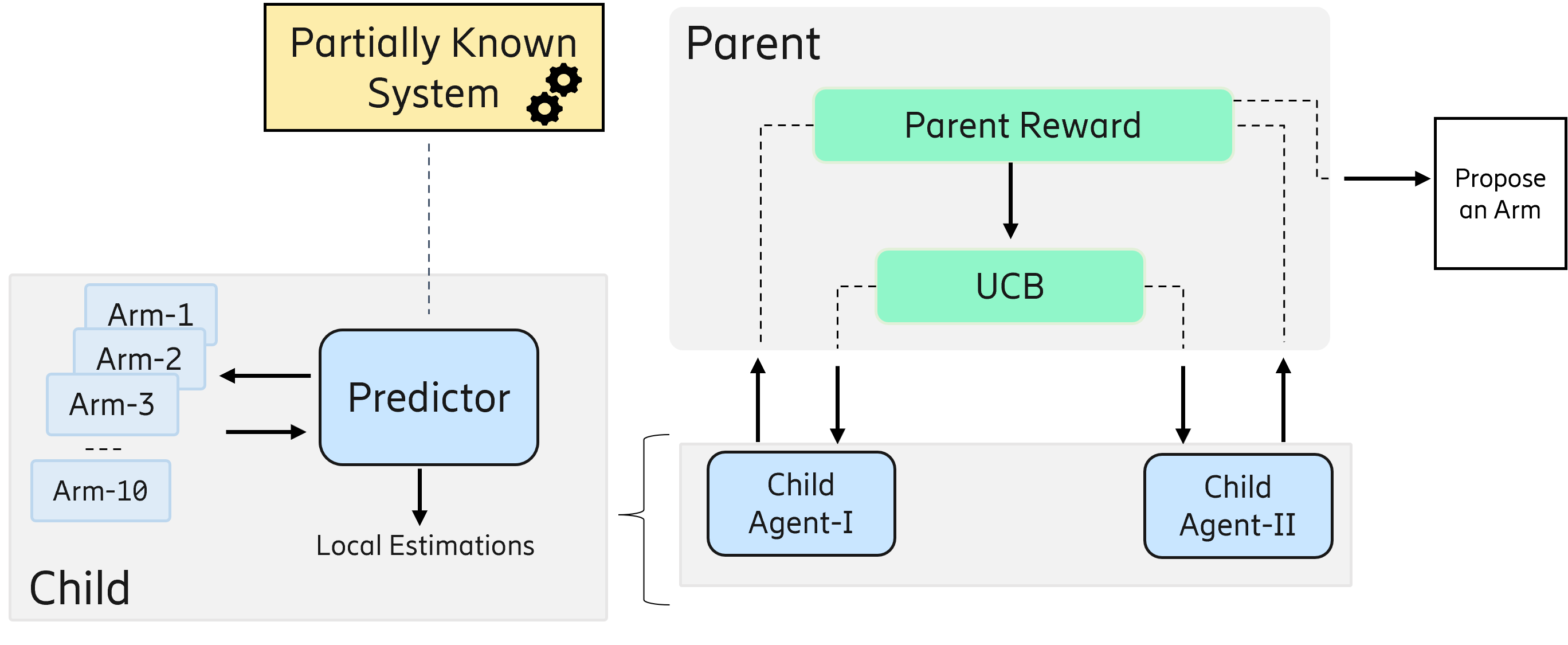}}
\caption{General structure of our approach for two service system case.}
\label{fig2}
\end{figure}

We note that with this approach, the actions from a CL can potentially impact the performance of other CL, which leads a conflict. In this paper, to address this challenge we propose to model the problem within a hierarchical  CLs framework, where the system consists of a parent CL and multiple child CLs. The parent CL has a more  holistic view of the network and  whereas the child CLs provide local estimations as a proxy for the parent agent. More specifically, each child CL is responsible for a particular network service KPI and is operating in a isolate mode and not aware of other child CLs. It is important for each CL to acquire new knowledge regarding its actions, and optimize its decision based on existing knowledge, which leads the well-known problem as trade-off between “exploration” and “exploitation” in the learning theory domain \cite{r23}. This hierarchical structure visualized in Figure \ref{fig2} allows for a more efficient exploration and exploitation of different network configurations in the dynamic network conditions.

In our MAB context, an \textit{arm} represents a network configuration aka action such as changing network priority level.
Let each child CL $i$ have $K$ arms, where $i \in \{1, 2, \ldots, N\}$, and $P_i^k$ denote the arm $k$ of child CL $i$,  $k \in \{1,2,..,K\}$.  The network aims to determine $P_i^{k,t}$ for each child CL $i$ at time $t$, considering the potential conflicts among the other CLs and dynamic network conditions. The objective is to maximize the expected cumulative reward over a time horizon $T$ with respect to network configurations, while minimizing the pseudo-regret, which quantifies the difference between the expected reward achieved by the optimal and selected arms.

We denote the reward of arm $k$ for CL $i$ at time $t$ as $X(P_i^{k,t})$. The rewards depend on the given target values for each service's KPI and are influenced by other uncontrolled random factors such a number of UE arriving or leaving the network, traffic load etc. The pseudo-regret, $R$, is defined as:
\begin{align}
R(T)= \mathbb{E}\left[\sum_{t=1}^{T}\sum_{i=1}^{N}[X(P_i^{k^*,t})-X(P_i^{\pi_i(t),t})]\right],
\label{eq0}
\end{align}
where $\pi_i(t)$ denotes our policy that  select the arm for CL $i$ at time $t$, and $k^*$ represents the optimal arm for CL $i$. We recall that our aim is to minimize this  while satisfying the KPI requirements of each intent under the given bandwidth limitation of the partially know network or the system. Next, we provide our solution for this problem in detail.

\section{Proposed Method}
We consider a hierarchical CL structure with which each child agent generates local estimations. Then, these estimation are sent to the parent agent to compute a global parent reward using a weighted sum of the local estimations. General architecture is depicted in Figure 1. 

Each child agent is associated to an intent KPI from a given service, and is responsible to fulfill the KPI requirement. For example, a child agent can be responsible to satisfy the QoE requirement of a video service. To do that, the child agent has a set of actions (i.e., arms) that it can take to allocate more resource for its KPI. Importantly, we take into account the fact that there can be many child agents and those agents must be act independently from each others. This is because handling multiple CLs with a single child agent cannot be tractable as it can lead to  a scalability issue in practice. In other words, when a child agent proposes an arm, it is not aware of what other child agents propose, and it considers the action of other agent as a random factor. The main responsibility of a child agent is to make local estimation on the impact of its arms on its KPI with the consideration of other factors such as number of UEs with the service and the arm of other child agents. 

The child agents are considered as low-level CLs that are governed by a high-level CL called a parent agent. When an action is proposed by a child agent, it is not directly applied to the real network but first it is evaluated at the parent agent. Parent agent is responsible in collecting local estimations from each child agent, then evaluate the situation with the consideration of total expected reward as well as the uncertainty that can be caused by the local estimations. This is because the local estimation of a child agent may not be within  a required confidence level which can be because of the fact that  some of the arms of a child CL  may not have been played often, causing high uncertainty. This uncertainty must be quantified for the parent agent to make an accurate decision. 
\subsection{Algorithm for Child CL} 
We develop Algorithm 1 with which each CL determines its proposal arm independently. Besides, child agents (CLs) are not allowed to propose every possible arm but the suggested set of arms, which we call active arms, coming from parent agent.
\begin{algorithm}[t]
\small
   \centering
   \caption{Child Agent-$i$}
   \label{alg:child}
\begin{algorithmic}[1]
   \STATE \textbf{Inputs:} {Information set at phase p:\\ $\tau(p)=(N, \gamma_{target})$,
   active arms for child-i:= $[K_i]$\\}
   \STATE \textbf{Parameters: }{$P_i = \{ P_{i,1},..., P_{i,K} \}$:= Arms for child agent-i\\}
   \hrulefill
   \FOR{each $k \in [K_i]$}
   \STATE {Pull arm k,}
   \STATE {Pick a random arm for another service; $P_{random}$\\}
   \STATE {By using reward predictor;\\}
    {\hspace*{1cm}$\mu_{k,n}^i=f_{predictor}(\tau(p),P_{i,k},P_{random})$}
   \STATE {Repeat step 5-6 n times and,\\ Calculate $\mu_k^i=\mathbb{E}[\mu_{k,n}^i]$} over these n trials
   \ENDFOR
   \STATE {Assign $\mu^i=[\mu_k^i,\forall k \in [K_i]]$}
   \STATE \textbf{return :} $\mu^i$
\end{algorithmic}
\end{algorithm}
\normalsize
At each phase $p$, there can be a new information denoted as $\tau(p)$ in which a KPI target or number of UEs with a service can change and parent agent provides a set of network configurations that child CLs can propose. Also, child agents takes its associated KPIs  with the target value as input (Line 1-2). As the system is dynamic, with changes in the number of UEs or targets in a given phase, child agents are informed of these changes while online learning is ongoing. Each child agent pulls arm from an active set determined denoted as $K_i$ by the parent agent  and predicts the reward of the proposed action (Line 3-7) by averaging several trials. We note that $\mu_{k,n}^i$ is the expected reward of child CL $i$ when it plays arm $k$ at trial $n$. At Line 5, the child agent takes what others can propose as a random factor due to blindness of the child agents. At Line 6, it uses a prediction model (i.e., a machine learning model) $f_{predictor}$ to predict the reward of the proposed arm $k$. $f_{predictor}$ takes the KPI target, arm of the child agents, and other network dynamics such as number of UEs as input, and it predicts the expected KPI value for the given inputs. We use a  multi-layer perception for our machine learning model.  This process happens multiple times by randomizing the arms from other agents, and the expected score is assigned to the proposed arm $k$ of the child agent (Line 9). This predictor is obtained based on partially observable system before learning process of system starts. The scores for each arm from the active set are returned in vector form to the parent agent which evaluates the effectiveness of the arm from each child agents.

\subsection{Multi arm bandit based Conflict Resolution (MABCR)} 

In our MABCR structure the operations of the parent agent is given in Algorithm 2,  and the parent agent acts as an evaluation module and makes decisions from an overall perspective.
\renewcommand{\algorithmicprint}{\textbf{break}}
\begin{algorithm}[t]
\small
   \centering
   \caption{Parent Agent}
   \label{alg:parent}
\begin{algorithmic}[1]
   \STATE \textbf{Input:} {N: number of child agents}
   \STATE \textbf{Initialize:} {Active arms in the parent, $A$:= $\{a_1,a_2,.., a_L\}$\\}
   {$a \in A$,\\}
   \STATE {$\gamma_{target}$:= Target KPIs for each service}
   \STATE {$N(a)$:= Number of times arm $a$ played,\\ $Q(a)$:= Parent reward with arm $a$}
   \STATE {Set $Q(a)$ $\leftarrow$ 0, $N(a)$ $\leftarrow$ 0, t $\leftarrow$ 0}
   \STATE {$w_{i}$:= importance of KPI $i$\\}
   \hrulefill
   \STATE{Denote $\tau(p)'$ as information set of previous phase}
   \WHILE{t $<$ end time}
   \STATE{Feed child agents with the new information, $\tau(p)$ at time t}
   \IF{ $\tau(p)' \neq \tau(p) $  (i.e., there is a target change,  $\gamma_{target} \neq \gamma_{target}'$}
        \STATE{Re-initialize the A as $\{a_1,a_2,.., a_L\}$}
   \ENDIF
   \STATE {Play each child agent and determine $\mu^i$ values from each child agent}
   \STATE {Calculate Q(a)$\leftarrow w_1\mu^1 + w_2\mu^2 + ... +w_N\mu^N , \forall a \in A$}
   \STATE {Update $N(a)\leftarrow N(a)+1, t \leftarrow t+1$}
   \STATE {Calculate UCB-1 Acquisition Function:\\}
   {\hspace*{1cm}$F(a,t) = Q(a)+\sqrt{\frac{2\ln{t}}{N(a)}}$}
   \STATE {Determine $a^* = argmax(F(a,t))$, \footnotesize \small}
   \IF{$a^*$ not in $A$}
        \STATE{Keep $a^*$at $A$}
   \ENDIF
   \IF{$\lvert A \rvert \neq 1$}
        \STATE{Eliminate $argmin(Q(a))$ from $A$}
   \ELSE
        \PRINT
   \ENDIF
   \IF{$t = t_{decision}$}
        \STATE{Propose an arm, $argmax(Q(a))$}
   \ENDIF
   \ENDWHILE
\end{algorithmic}
\end{algorithm}
\normalsize

Initially all possible combinations of child arms are considered as potential final arm that can be applied to the real network. The set $A$ denotes the active set, and an element of set $A$, $a$,  is called as parent arm which is one unique combination of child arms. For example, in the case where there are only two child agents with each having $K$ arms, $a$ is a vector including ($P_{1}^k, P_{2}^l$) where $P_{1}^k$ and $P_{2}^l$ are the particular arms of child agent 1 and 2 respectively where $k,l \in \{1,2,..,K\}$. $Q(a)$ and $N(a)$ are the reward and number of times the arm $a$ played up to time $t$, which are initially set to zero. Also, Algorithm 2 takes the number of child agent $N$, possible active set of arms $A$, the KPI targets of services $\gamma_{target}$.
Alg. 2 first checks if  there is a new information at a new phase denoted by $\tau(p)$. At this new phase $\tau(p)$, as an example, a KPI target or number of UEs with a service can change and hence we check whether if there is such a change at Line 7-10. If there is a change, the active set is re-initialized, and reset to all possible combination of child arms (Line 11). Parent agent then starts to collect the expected reward from child agents with Algorithm 1 (Line 13). Then the parent agent calculates the weighted average reward of each arm $a$ (Line 14) and $N(a)$ is updated. We note that $w_i$ is the weight of intent expectation $i$. At Line 16, by using an Upper Confidence Bound (UCB) algorithm, the arm that maximizes $F(a,t)$ is selected and added to the set $A$. We note that $F(a,t)$ does not only take into account the average reward but also the uncertainty with the arm $a$. Until the parent agents makes a final decision on the arm to be applied to real network (Line 26), the process continues and the set $A$ is generated. Also, the arm with minimum reward is removed in the active set (Line 22). The method we are using to create active set over eliminations is based Fed2-UCB algorithm given in \cite{fmab}. Alg. 2 keeps the arms that maximizes reward and minimizes the uncertainty, and it eliminates the arms with minimum reward. At final stage, the only the arms that manage to be kept at the set $A$ are sent back to the child agents.

This elimination process continues until there is only one element remaining in the active set (as described in Lines 21-22) and at the decision point $t_{decision}$, parent agent selects the arm in the set $A$ that maximizes the weighted average reward (Line 26-27). Ultimately, the parent CL continuously proposes the best possible arm based on the current network configurations, while the online algorithm strives to converge to the optimal set of arms.

\section{Numerical Analyses}
\subsection{Simulation Environment}
We carried out simulation experiments to assess our proposal and benchmark it against different baseline algorithms. The simulation environment, designed in Python, is based on a in-house network emulator. The network emulator consists of containerized entities including network functions, UEs, and application instances. These individual components are implemented as web servers providing a set of APIs enabling configuration management. Based on these capabilities, the set of actions (arm) that can be executed to reconfigure the network is as follows: 

\textit{Priority:} Changing the priority of a service instance that reconfigures  how much  resource a service can take.

\textit{MBR (Maximum Bit Rate):} Changing the maximum throughput that can be achieved by a particular UE. 

We implemented two different service types in the network emulator that are: (1) Conversational video and (2) Ultra-reliable low latency communication (URLLC). For the URLLC service, the primary KPIs are latency and reliability (i.e., packet loss). Conversely, conversational video represents an enhanced mobile broadband (eMBB) service, which is not highly sensitive to latency. Instead, this type of service is evaluated using Key Quality Indicators (KQIs) in addition to KPIs to assess user-perceived performance. Therefore, we define Quality of Experience (QoE) as the target KPI, based on both network and application layer metrics. 

We first run our in-house network emulator under different configurations and collected realistic data from the emulator, and create a large data set including the actions,  the throughput, latency, QoE values for the UEs of different services. First we observed that there is a good correlation between the achieved throughput and the target services KPIs, QoE and latency. Hence, 
one can estimate the target KPIs from the achieved throughput. To do that,  by using this data set, we next attempt to learn the functional relations between latency and the achieved throughput  for URLLC service denoted as $f_u(th)$, where $th$ is the achieved throughput by a UE of URLLC service and $f_u(th)$ represents the latency of the URLLC UE for a given throughput $th$.  Also,  we find the function $f_v(th)$ representing the relation between QoE and the achieved throughput for video service. Those functions are derived by  using simple polynomial approximation by using the data set as the relations are not linear. Once we learn these functions and the achieved throughput, one can easily estimate the KPI values.  Later, those functions  are used in our simplified simulation environment where we test our approach in the rest of this paper, and utilize these functions to learn how much throughput each service needs to achieve to satisfy their KPI requirements such as latency and QoE. 

In our simplified environment developed in Python, the throughput of  a UE with service $s$ is calculated as follows:
\begin{align}
th_s = \min \{B \cdot \frac{P_S}{\sum P_S}, MBR\}
\end{align}
where $B$ is the total system bandwidth. We set $B$ to 10 Mbps, and $MBR$ to 3 Mbps for all UEs in our setup. Also, $P_s$ is the priority level of service $s$. In order to be more realistic and also to understand how efficient our proposal is in terms of resource allocation, we also implemented a physical layer model where Physical Resource Blocks (PRBs) are determined based on channel condition. Specifically, for each UE, the Quality Class Identifier (QCI) parameters of channels are randomly generated within a range of 0 to 15 and mapped to the corresponding Modulation and Coding Scheme (MCS) and spectral efficiency values as outlined in Table \ref{table:MCS}. This table presents the Channel Quality Indicator (CQI), MCS index, and spectral efficiency values in 5G New Radio (NR) when the number of layers is two. Under the given channel conditions, the amount of PRBs required can be calculated as \cite{3gpp}:
\begin{align}
N_{PRB}=\frac{th}{180 \text{kHz}/\text{RB} \cdot \text{SE}_i},\label{eq2}
\end{align}
where $th$ represents the achieved throughput for the UE of a service, and  RB represents an adjacent group of 12 subcarriers, where each subcarrier has a bandwidth of 15 KHz, and $SE_i$ is the spectral efficiency value of the i-th user calculated based on Table \ref{table:MCS}.
\begin{table}[tbp]
\caption{Modulation and Coding Scheme Table, 38.214 3GPP Technical Report \cite{3gpp}}\label{table:MCS}
\begin{center}
\begin{tabular}{|c|c|c|}
\cline{1-3} 
\textbf{\textit{QCI}}&\textbf{\textit{MCS}}&\textbf{\textit{SE}}\\
\hline
1&0&0.1523\\
2&1&0.2344\\
3&2&0.3770\\
4&3&0.6016\\
5&4&0.8870\\
6&5&1.1758\\
7&6&1.4766\\
8&7&1.9141\\
9&8&2.4063\\
10&9&2.7305\\
11&9&3.3223\\
12&9&3.9023\\
13&9&4.5234\\
14&9&5.1152\\
15&9&5.5547\\
\hline
\end{tabular}
\end{center}
\end{table}

In summary, the control action (i.e., arm) of each services or CL is the service priority levels dynamically determined by our MABCR online algorithm. First, our approach using Algorithm 1 and 2 provides the priority levels, $P_s$, for the services  (MBR are fixed for all UEs), then we  find the throughput, $th$, of the services with  the suggested  priority levels by our MABCR according to eq. (4). Then, by using the learnt functions $f_u(th)$ and $f_v(th)$, we  determine the achieved KPI levels (latency and QoE). Moreover, by using eq. (5),  we decide on how many PRBs are needed to achieve the required throughput level. Later, we asses the efficiency of our solution based on the number of the allocated PRBs. One of our future aim is to make the complete implement and full integration of our approach to our in-house network emulator.

\begin{table}[tbp]
\caption{Scenario-I}
\begin{center}
\begin{tabular}{|c|c|c|c|c|c|}
\hline
\multicolumn{1}{|c|}{\textbf{Phase}}&\multicolumn{5}{|c|}{\textbf{Parameters}}\\
\cline{1-6} 
\textbf{\textit{}}&\textbf{\textit{N-I}}&\textbf{\textit{N-II}}& \textbf{\textit{T}}&\textbf{\textit{$\gamma$}}&\textbf{\textit{$\eta$}}\\
\hline
I&5&5&20&3.4&60\\
II&5&5&20&3.45&60\\
III&5&5&20&3.5&60\\
IV&5&5&20&3.55&60\\
V&5&5&20&3.6&60\\
\hline
\multicolumn{5}{l}{N-I,II: Number of UEs for the services}
\end{tabular}
\label{tab3}
\end{center}
\end{table}
\subsection{Simulation Scenarios}
Our experiments compare the performance of three approaches: i-) our MABCR online learning method with UCB, ii-) the greedy child's approach, and iii-)  the MABCR online learning with Thompson sampling (TS). The \textit{greedy child's policy} is another version of our MABCR-Online Learning without a parent/evaluator modules. Hence, each child aims to achieve its target greedily and without a holistic control mechanism. The MABCR online learning with Thompson sampling is similar to our proposed approach  except the acquisition function \cite{ts}. Thompson sampling selects each arm randomly based on its probability of being the optimal choice. On the other hand, UCB in MABCR online learning is designed to find a balance between the most beneficial arms and taking uncertain actions that could potentially have a high reward.
\begin{table}[tbp]
\caption{Scenario-II}
\begin{center}
\begin{tabular}{|c|c|c|c|c|c|}
\hline
\multicolumn{1}{|c|}{\textbf{Phase}}&\multicolumn{5}{|c|}{\textbf{Parameters}}\\
\cline{1-6} 
\textbf{\textit{}}&\textbf{\textit{N-I}}&\textbf{\textit{N-II}}& \textbf{\textit{T}}&\textbf{\textit{$\gamma$}}&\textbf{\textit{$\eta$}}\\
\hline
I&5&5&20&3.4&60\\
II&5&5&20&3.4&55\\
III&5&5&20&3.4&50\\
IV&5&5&20&3.4&45\\
V&5&5&20&3.4&40\\
\hline
\end{tabular}
\label{tab4}
\end{center}
\end{table}
\begin{table}[]
\caption{Scenario-III}
\begin{center}
\begin{tabular}{|c|c|c|c|c|c|}
\hline
\multicolumn{1}{|c|}{\textbf{Phase}}&\multicolumn{5}{|c|}{\textbf{Parameters}}\\
\cline{1-6} 
\textbf{\textit{}}&\textbf{\textit{N-I}}&\textbf{\textit{N-II}}& \textbf{\textit{T}}&\textbf{\textit{$\gamma$}}&\textbf{\textit{$\eta$}}\\
\hline
I&5&5&10&3.35&40\\
II&5&5&10&3.35&45\\
III&5&5&10&3.4&50\\
IV&5&5&10&3.45&55\\
V&5&5&10&3.6&55\\
VI&5&5&10&3.6&55\\
VII&4&5&10&3.6&55\\
VIII&3&5&10&3.4&55\\
IX&2&5&10&3.5&50\\
X&1&5&10&3.6&55\\
\hline
\end{tabular}
\label{tab5}
\end{center}
\end{table}

\begin{figure*}[!ht]
\centering
\subfloat[Latency - SLA Violations]{\includegraphics[width=0.3\linewidth]{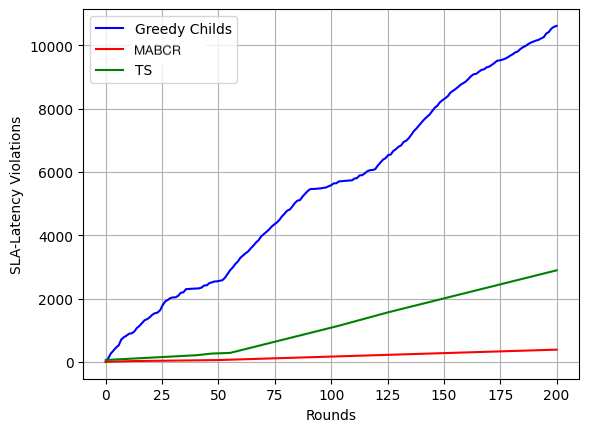}%
\label{system:first_graph}}
\hfil
\subfloat[QoE - SLA Violations]{\includegraphics[width=0.3\linewidth]{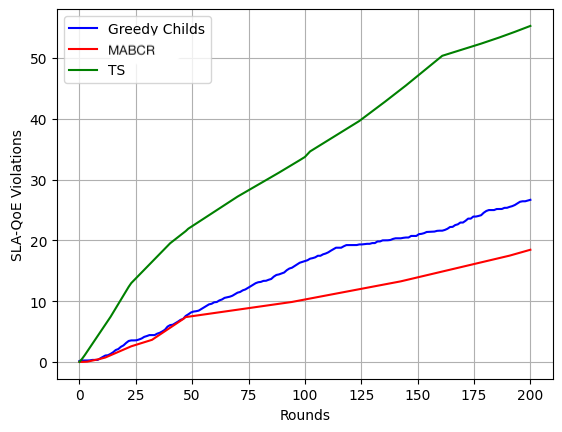}%
\label{system:second_graph}}
\hfil
\subfloat[PRB Allocations]{\includegraphics[width=0.3\linewidth]{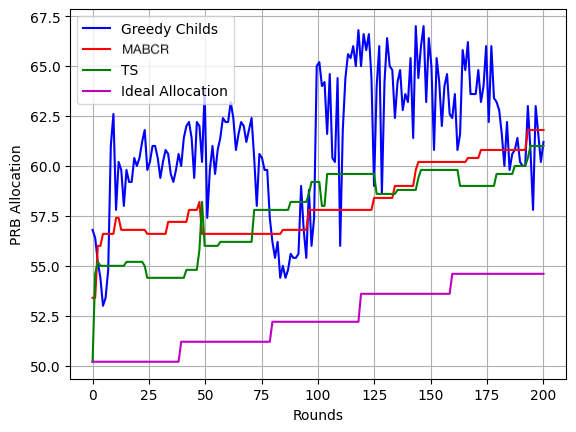}%
\label{system:second_graph}}
\caption{Each plot above shows the results for scenario given Table-I  with confidence interval 95$\%$ for the Scenario-I.}
\label{fig:scenario1}
\end{figure*}

\begin{figure*}[!ht]
\centering
\subfloat[Latency - SLA Violations]{\includegraphics[width=0.3\linewidth]{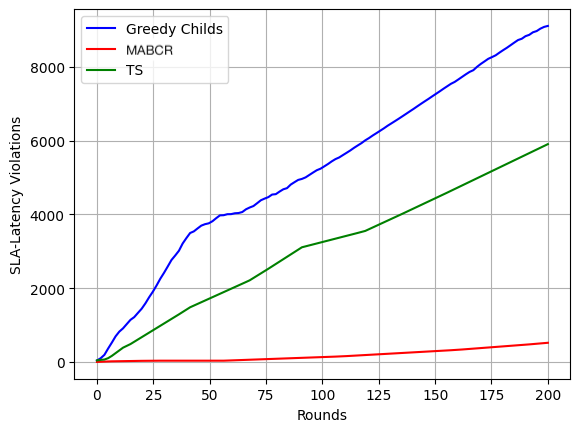}%
\label{system:first_graph}}
\hfil
\subfloat[QoE - SLA Violations]{\includegraphics[width=0.3\linewidth]{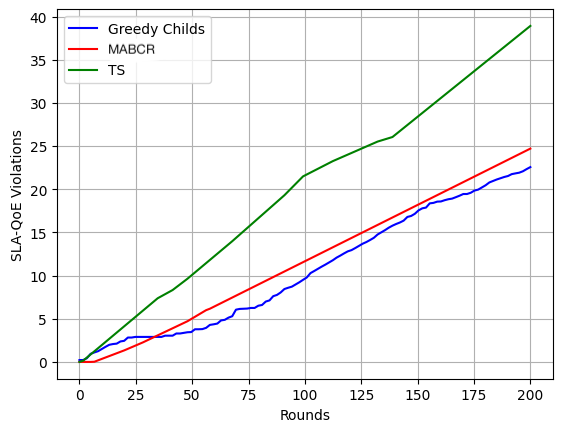}%
\label{system:second_graph}}
\hfil
\subfloat[PRB Allocations]{\includegraphics[width=0.3\linewidth]{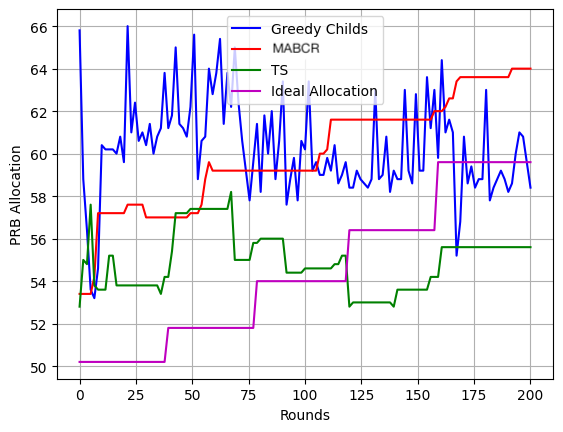}%
\label{system:second_graph}}
\caption{Each plot above shows the results for scenario given Table-II  with confidence interval 95$\%$ for the Scenario-II}
\label{fig:scenario2}
\end{figure*}

\begin{figure*}[!ht]
\centering
\subfloat[Latency - SLA Violations]{\includegraphics[width=0.3\linewidth]{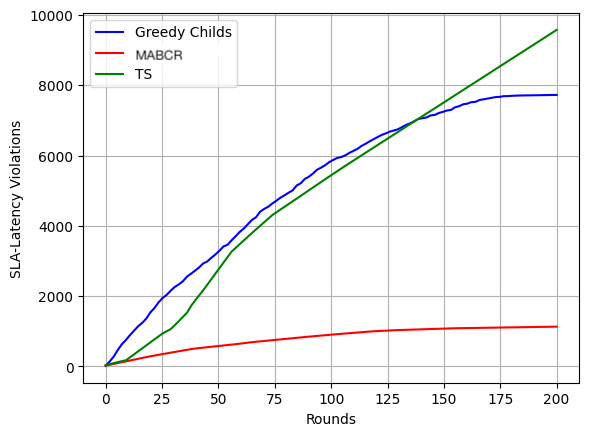}%
\label{system:first_graph}}
\hfil
\subfloat[QoE - SLA Violations]{\includegraphics[width=0.3\linewidth]{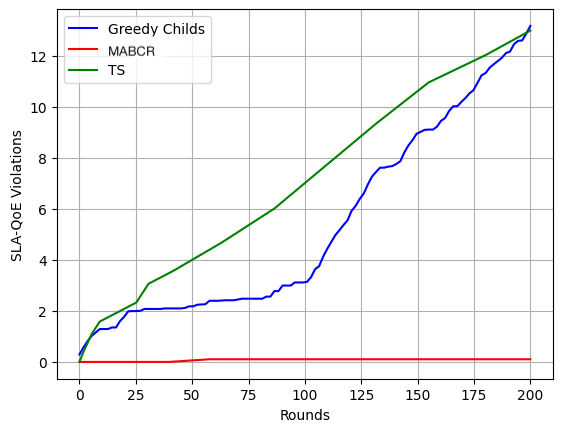}%
\label{system:second_graph}}
\hfil
\subfloat[PRB Allocations]{\includegraphics[width=0.3\linewidth]{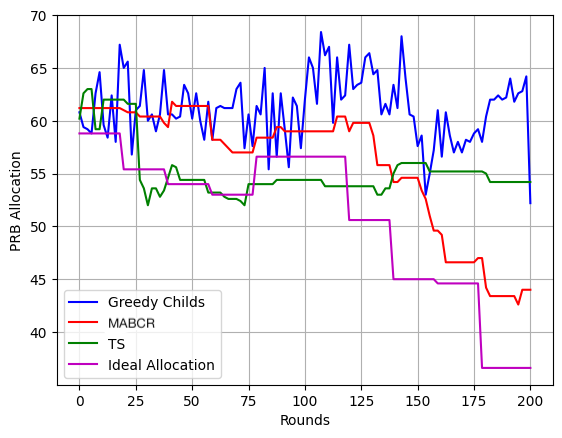}%
\label{system:second_graph}}
\caption{Each plot above shows the results for scenario given Table-III  with confidence interval 95$\%$ for the Scenario-III.}
\label{fig:scenario3}
\end{figure*}
We assume that there are intents containing the KPI expectation of video and URLLC services. For example an intent of URLLC service may indicate an expectation on latency, i.e., latency should be less than 30 ms. Importantly, the intent expectations can vary over time and the network is supposed to adapt these variations. To evaluate the performance of the algorithms, two key measures are used: the total number of SLA violations (deviation from intent expectations) and the amount of allocated PRBs. The cumulative number of SLA violations is calculated by adding up all deviations from the service targets. To effectively quantify the performance of our approach, we create  three different dynamic scenarios where the  number of UEs with both services and their associated QoE and latency targets vary over time.  These scenarios are outlined in Table \ref{tab3}, \ref{tab4}, and \ref{tab5}, respectively. The total bandwidth, $B$, is set to 10 Mbps and MBR of each UE is set to 3 Mbps.

Specifically, in scenario I given in Table \ref{tab3} where $\gamma$ and $\eta$ stands for QoE and latency targets respectively, and there are 5 UEs for both video and URLLC services and delay requirement for URLLC service is set to 60 ms. However, QoE expectation (i.e., target) of each UE of the video services varies at every 20 seconds. For Scenario 2 described in Table \ref{tab4},  we now fix the QoE target of video service to 3.4 and changes the delay requirement of URLLC service between 60 and 40 ms. For scenario I and II, we have five phases where at each phase we have a different QoE or delay requirements, and each phase takes 20 seconds in the simulation. At the last scenario with ten phases with 10 seconds of simulation time and  given in Table \ref{tab5}, we take a more challenging case where both QoE and delay requirements of video and URLLC services vary over time, respectively.

The horizontal axis in the figures represents the algorithm's decision stages. The duration parameter in the table is determined based on the total simulation time of 500 loop duration. Multiple sample are conducted to obtain an ensemble average to ensure reliable results and mitigate the impact of randomness. A t-test is used to generate a 95\% confidence interval for the cumulative SLA violations graphs.

\subsection{Results}

As illustrated in Figure \ref{fig:scenario1}, the algorithms are assessed in terms of QoE changes, which become increasingly challenging with each round. Here, achieving latency targets appears more feasible compared to satisfying QoE targets simultaneously. As a result, the algorithms must compromise on the attainable targets to manage this trade-off effectively. Our proposed algorithm shows the most efficient resolution for this trade-off, managing the QoE satisfaction and latency targets in an adaptive way.

The TS approach seeks to find a compromise between QoE and latency targets. However, as evident from the results in Figure \ref{fig:scenario1}, it falls short of achieving a good solution as our proposed algorithm. On the other hand, the greedy approach tends to deviate more compared to the other methods. Notably, greedy child's focus solely on one target, giving it full importance without considering the trade-off between services. From a single service's perspective, the greedy approach might appear more successful than TS. However, when evaluating the overall cell of the network, our proposed algorithm outperforms both the greedy child's and TS approaches in handling the trade-off between QoE and latency targets.

When the targets change, the algorithm adjusts the arms accordingly. In some cases, they may continue with the same arms, incurring minor violations, while in other instances, they may modify their arms slightly after target alterations. However, the exploration-exploitation dilemma faced by online learning algorithms introduces some perturbations in arm selection. The parent, serving as an evaluation module, plays a stabilizing role by considering all system circumstances, allowing the arms to remain consistent for consecutive rounds. As a result, the allocation of PRBs remains more stable in our learning algorithms with parents, in contrast to the greedy child case. Without the evaluator module, we observe a significant level of fluctuation caused by unstable decisions,  hindering the discovery of mutually beneficial solutions for our resource allocation problem, as evident in the Figures showing PRB allocation .

Figure \ref{fig:scenario1}, \ref{fig:scenario2} and \ref{fig:scenario3} also show that the PRB allocation of the greedy approach fluctuates, causing excessive resource occupation. Our proposed algorithm's allocation has a small margin compared to the ideal allocation, as seen in Figure  \ref{fig:scenario1}, \ref{fig:scenario2} and \ref{fig:scenario3}. However, in some cases, the TS approach may deviate from the ideal allocation, resulting in violations due to insufficient resource deployment. The margin between these algorithms and the ideal allocation is mainly due to the granularity of bandwidth sharing for each service. Moreover, we add a 0.5 Mbps margin to the ideally required total bandwidth for each phase, and we convey this information to the parent  to ensure more realistic constraints. 

The greedy child's approach behaves selfishly, attempting to satisfy all service level agreements without any authority like a parent or evaluator. On the other hand, the results for the MABCR without UCB approach demonstrate the need for an acquisition function to explore the environment and resolve conflicts optimally. Our proposed method offers a comprehensive approach that addresses conflicts between service targets and resource allocation, particularly in terms of PRB allocation.

\section{Conclusions}
Our work has shown that components of our proposed approach within the hierarchical structure can efficiently work together in an online learning process to meet the intent expectations. MAB framework in a hierarchical structure of CLs, comprised of a parent CL and multitude of child CLs, is in charge of determining the network configuration, while each child CL focuses on prediction on network services thanks to partially known system. Our extensive simulation results show that our scheme can adapt the dynamically changing intent expectation and environment and also takes into account the resource efficiency. Next we would like to implement our idea directly within our in --house emulator after  new features are added. Also, it would be interesting to have different types of network configurations that can be proposed by the agents. 
\balance
\nocite{*}


\end{document}